\documentclass[prd,aps,nofootinbib,notitlepage,showpacs,preprintnumbers]
{revtex4-1}
\usepackage{graphicx,epsf,amsmath,amsfonts,amssymb,amsbsy}
\usepackage{epsfig}
\newcommand{\ds}{\displaystyle}
\newcommand{\vev}[1]{\langle#1\rangle}
\newcommand{\mat}{\left ( \begin{array}}
\newcommand{\emat}{\end{array} \right )}
\newcommand{\vect}{\left ( \begin{array}{c}}
\newcommand{\evect}{\end{array} \right )}

\begin{document}

\hfill HU-EP-11/11

\title{Chiral density waves in the
NJL$_2$ model with quark number and isospin chemical potentials}
\author{D.~Ebert$^{1)}$, N.V. Gubina$^{2)}$, K.G. Klimenko$^{3)}$,
  S.G. Kurbanov$^{2)}$ and V.Ch. Zhukovsky$^{2)}$}
\affiliation{$^{1)}$ Institute of Physics, Humboldt-University
Berlin, 12489 Berlin, Germany} \affiliation{$^{2)}$ Faculty of
Physics, Moscow State University, 119991, Moscow, Russia}
\affiliation{$^{3)}$ IHEP and University "Dubna" (Protvino branch),
142281, Protvino, Moscow Region, Russia}

\begin{abstract}
We investigate the phase portrait of the (1+1)-dimensional massless
two-flavored NJL$_2$ model containing a quark number chemical
potential $\mu$ and an isospin chemical potential $\mu_I$ in the
limit of a large number of colors  $N_c\to\infty$. Particular
attention is paid to the question to what extent the inclusion of an
isospin asymmetry affects chiral condensates to have a spatial
inhomogeneity in the form of the so-called chiral density waves
(CDW) (chiral spirals). It is shown that at zero temperature and
comparatively small values of $\mu$, i.e. at $\mu<\mu_c\approx
0.68M_0$ ($M_0$ is the dynamical quark mass in the vacuum) only the
homogeneous charged pion condensation phase is realized for
arbitrary nonzero values of $\mu_I$. Contrary to this, for large
values of $\mu>\mu_c$, two CDW phases appear in the
$(\mu_I,\mu)$-phase diagram of the model.  In the first phase, CDWs
are clockwise twisted chiral spirals and in the second phase they
are counterclockwise. The influence of nonzero temperature on the
formation of the CDW phases is also investigated.
\end{abstract}


\maketitle
\section{Introduction}

During the last decade, much attention has been attracted to the
investigation of the QCD phase diagram in terms of quark number as
well as isospin chemical potentials. First of all, this is motivated
by heavy-ion collision experiments where dense baryonic matter has
an evident isospin asymmetry, i.e. different neutron and proton
contents of initial ions. Moreover, the dense hadronic/quark  matter
inside compact stars is also expected to be isotopically asymmetric.
Generally speaking, it is understood that one of the important
challenges for QCD  is to describe the dense and hot baryonic matter
in different physical situations. However, in the above mentioned
realistic situations the quark density is rather small, and weak
coupling QCD analysis is not applicable. So, different
nonperturbative methods or effective theories such as chiral
effective Lagrangians and especially  Nambu -- Jona-Lasinio (NJL)
type models \cite{njl} are usually employed for the consideration of
the properties of dense and hot baryonic matter under the conditions
of heavy-ion experiments or in the compact stars interior, i.e. in
the presence of external factors such as temperature, chemical
potentials, magnetic field, finite size effects, etc. (see, e.g.,
\cite{alford,hiller,klim,warringa,incera,ebert,vshivtsev} and
references therein). In particular, phenomena of dense quark matter
like color superconductivity \cite{alford,klim,warringa,incera}, as
well as charged pion condensation \cite{son,ek,abuki}  were
investigated in the framework of these QCD-like effective models.

It should be noted that an effective description of QCD in terms of
NJL models, i.e. through an employment of four-fermionic theories in
(3+1)-dimensional spacetime, is usually valid only at {\it
comparatively low} energies and densities. At the same time,
(1+1)-dimensional Gross-Neveu (GN) type models \cite{gn,ft} are
valid also at high energy and density, and due to their properties
of renormalizability, asymptotic freedom and spontaneous chiral
symmetry breaking, can also be used for a reasonable qualitative
modeling of QCD even at finite temperature and hadron density
\cite{wolff,barducci,chodos,ohwa,thies}.
 Due to the relative simplicity of GN models in the
leading order of the large $N_c$-expansion ($N_c$ is the number of
colored quarks), their use is convenient for the application of
nonperturbative methods in quantum field theory \cite{okopinska}.
Moreover, it is worth noting that it is in the leading order of the
large $N_c$-expansion that the well known no-go theorem of
Mermin--Wagner--Coleman \cite{coleman}, apparently forbidding the
spontaneous breaking of continuous symmetries in the
(1+1)-dimensional models, becomes invalid
\cite{barducci,chodos,ohwa,thies}. (It means that in the large $N_c$
limit quantum fluctuations, which would otherwise destroy a
long-range order corresponding to a spontaneous symmetry breaking,
are suppressed by $1/N_c$ factors.) Note also that GN type models
are quite suitable for the description of physics in quasi
one-dimensional condensed matter systems such as polyacetylene
\cite{caldas}.

Thus, such phenomena of dense QCD as color superconductivity, where
the color group is broken spontaneously, and charged pion
condensation, where spontaneous breaking of the continuous isospin
symmetry takes place, might be modeled in terms of renormalizable
(1+1)-dimensional GN type models (see, e.g., \cite{chodos,ohwa} and
\cite{ektz,ek3,ek2}, respectively).

In our previous papers \cite{ektz,ek3,ek2} the phase diagram of a
(1+1)-dimensional $SU_L(2)\times SU_R(2)$ symmetric NJL model
\footnote{In this paper we shall use the notation NJL model for
theories with four-fermionic interactions also for (1+1)-dimensional
models with {\it a continuous chiral symmetry group} instead of
``chiral 2D GN model'' due to the fact that the chiral structure of
the Lagrangian is indeed closely related to the (3+1)-dimensional
NJL model.} with two massless or massive quark flavors was
investigated in the leading order of the $1/N_c$-expansion and in
the presence of the quark number- as well as isospin chemical
potentials. There we have considered the case of order parameters
(condensates) that are homogeneous, i.e. independent of the space
coordinate. The situation corresponds to the conserved Lorentz and
spatial translational invariance and is adequate to physical systems
in vacuum, i.e. at zero chemical potentials. In dense baryonic
matter, i.e. at nonzero quark number chemical potential, there might
appear new phases with a spatially inhomogeneous chiral condensate
which destroys both chiral and spatial translational invariance of
the system (see the relevant discussions made in the framework of
both (1+1)-dimensional \cite{thies,thies2,misha} and
(3+1)-dimensional \cite{3+1,nakano,nickel,maedan,zfk} models). Thus,
in this paper and in contrast to \cite{ektz,ek3,ek2},  we consider
the phase portrait of the above mentioned massless $SU_L(2)\times
SU_R(2)$ symmetric NJL model with two chemical potentials in the
leading order of the $1/N_c$-expansion taking into account the
possibility that the chiral condensate might become inhomogeneous
and take the form of a (dual)chiral density wave (CDW). In this case
the scalar quark-antiquark condensate, $\vev{\bar qq}$, and the
pseudoscalar condensate of neutral $\pi^0$ mesons, $\vev{\bar
q\gamma^5\tau_3q}$, form a chiral spiral, i.e.
\begin{eqnarray*}
\vev{\bar qq}\sim \cos 2bx,~~~~~\vev{\bar q\gamma^5\tau_3q}\sim~\sin
2bx,
\end{eqnarray*}
where $\tau_3$ is the isospin Pauli matrix, $x$ is the space
coordinate, and $b$ is a wave vector which has to be determined
dynamically through the thermodynamic potential. It is necessary to
point out that the inhomogeneous CDW-condensate is relevant to dense
quark matter \cite{pisarski} and the chiral magnetic effect
\cite{basar}. Both phenomena probably might be observed in heavy-ion
collision experiments, where isotopic asymmetry is an inevitable
property. So, we believe that the investigation of CDW condensates
in the framework of the two-dimensional NJL model with isospin
chemical potential could shed some light on  physics of heavy-ion
collisions. It should be noted, however, that our investigation of a
condensate inhomogeneity in the form of the chiral density wave is
only a first step. There may exist, at least at zero isospin
chemical potential, other more preferable spatially non-uniform
ground state configurations of the chiral condensate like, for
instance, chiral crystals \cite{thies2,misha,nickel}) (in the last
case only the scalar $\vev{\bar qq}$ condensate is an inhomogeneous
quantity, and other condensates are homogeneous ones), but, in
general, they are much harder to deal with. For technical reasons,
in studying CDW configurations we do not take into account a nonzero
bare (current) quark mass, although recently some efforts to get rid
of this assumption have been made \cite{maedan}.

The paper is organized as follows. In Section II, we derive in the
leading order of the large $N_c$-expansion the expression for the
thermodynamic potential of the $SU_L(2)\times SU_R(2)$ symmetric
massless NJL$_2$ model with quark number chemical potential $\mu$
and isospin chemical potential $\mu_I$ at zero temperature. Here we
also consider the possibility of a spatial inhomogeneity for the
chiral condensates in the form of the so-called chiral density
waves. First, the phase portrait of the model is discussed in the
simple case of spatially homogeneous condensates in Section III, and
then, in Section IV, the phase structure of the model in terms of
$\mu$ and $\mu_I$ and at zero temperature is investigated for CDW
inhomogeneous phases. The influence of nonzero temperature on the
formation of CDW phases is considered in Section V. Finally, Section
VI presents some concluding remarks.

\section{ The model and its effective action}
\label{effaction}

We consider a (1+1)-dimensional NJL-type model with two massless
quark flavors ($u$ and $d$ quarks) to mimic properties of real dense
quark matter.  Its Lagrangian has the form
\begin{eqnarray}
&&  {\cal L}=\bar q\Big [\gamma^\rho\mathrm{i}\partial_\rho
+\mu\gamma^0+\frac{\mu_I}2 \tau_3\gamma^0\Big ]q+ \frac {G}{N_c}\Big
[(\bar qq)^2+(\bar q\mathrm{i}\gamma^5\vec\tau q)^2 \Big ],
\label{1}
\end{eqnarray}
where our choice for the gamma matrices in (1+1)-dimensions is as
follows:
$\gamma^0=\sigma_2, \gamma^1=\mathrm{i}\sigma_1,
\gamma^5=\gamma^0\gamma^1=\sigma_3$, and the quark field $q(x)\equiv
q_{i\alpha}(x)$ is a flavor doublet ($i=1,2$ or $i=u,d$) with
corresponding  Pauli matrices $\tau_k$ ($k=1,2,3$) and color
$N_c$-plet ($\alpha=1,...,N_c$), as well as a two-component Dirac
spinor (the summation in (\ref{1}) over flavor, color, and spinor
indices is implied). The quark number chemical potential $\mu$ in
(\ref{1}) is responsible for the nonzero baryonic density of quark
matter, whereas the isospin chemical potential $\mu_I$ is taken into
account in order to study asymmetric quark matter at nonzero isospin
densities (in this case the densities of $u$ and $d$ quarks are
different).  Evidently, the model (\ref{1}) is a generalization of
the original (1+1)-dimensional Gross-Neveu model \cite{gn} with a
single  quark to the case of two quark flavors and additional
chemical potentials. As a result, we have for our model a more
complicated chiral symmetry group. Indeed, at
$\mu_I =0$ apart from the global color SU($N_c$) symmetry, the
Lagrangian (\ref{1}) is invariant under transformations of the
chiral $SU_L(2)\times SU_R(2)$ group. However, at $\mu_I \ne 0$ this
symmetry is reduced to $U_{I_3L}(1)\times U_{I_3R}(1)$, where
$I_3=\tau_3/2$ is the third component of the isospin operator (as
usual the subscripts $L,R$ mean that the corresponding group
acts only on the left, right handed spinors, respectively).
Evidently, this symmetry can also be presented as $U_{I_3}(1)\times
U_{AI_3}(1)$, where $U_{I_3}(1)$, $U_{AI_3}(1)$ denote the isospin
and the axial isospin subgroups respectively. Quarks are transformed
under these subgroups as $q\to\exp (\mathrm{i}\alpha\tau_3) q$ and
$q\to\exp (\mathrm{i} \alpha\gamma^5\tau_3) q$, respectively
\footnote{\label{f1,1} Recall that~~ $\exp
(\mathrm{i}\alpha\tau_3)=\cos\alpha
+\mathrm{i}\tau_3\sin\alpha$,~~~~ $\exp
(\mathrm{i}\alpha\gamma^5\tau_3)=\cos\alpha
+\mathrm{i}\gamma^5\tau_3\sin\alpha$.}.
Notice that Lagrangian (\ref{1}) is parity invariant.

The linearized version of the Lagrangian (\ref{1}), which contains
composite bosonic fields $\sigma (x)$ and $\pi_a (x)$ $(a=1,2,3)$,
has the following form:
\begin{eqnarray}
\tilde {\cal L}\ds &=&\bar q\Big [\gamma^\rho\mathrm{i}\partial_\rho
+\mu\gamma^0+ \nu\tau_3\gamma^0-\sigma
-\mathrm{i}\gamma^5\pi_a\tau_a\Big ]q
 -\frac{N_c}{4G}\Big [\sigma^2+\pi_a^2\Big ],
\label{2}
\end{eqnarray}
where $\nu=\mu_I/2$. Evidently, the Lagrangian (\ref{2}) is
equivalent to (\ref{1}), which simply follows from the use of the
following constraint equations for the bosonic fields
\begin{eqnarray}
\sigma(x)=-2\frac G{N_c}(\bar qq);~~~\pi_a (x)=-2\frac G{N_c}(\bar q
\mathrm{i}\gamma^5\tau_a q). \label{200}
\end{eqnarray}
Furthermore, it is clear from (\ref{200}) and footnote \ref{f1,1}
that the bosonic fields transform under the isospin $U_{I_3}(1)$ and
axial isospin $U_{AI_3}(1)$ subgroups in the following manner:
\begin{eqnarray}
U_{I_3}(1):~~&&\sigma\to\sigma;~~\pi_3\to\pi_3;~~\pi_1\to\cos
(2\alpha)\pi_1+\sin (2\alpha)\pi_2;~~\pi_2\to\cos (2\alpha)\pi_2-\sin
(2\alpha)\pi_1,\nonumber\\
U_{AI_3}(1):~~&&\pi_1\to\pi_1;~~\pi_2\to\pi_2;~~\sigma\to\cos
(2\alpha)\sigma+\sin (2\alpha)\pi_3;~~\pi_3\to\cos
(2\alpha)\pi_3-\sin (2\alpha)\sigma.
\label{201}
\end{eqnarray}
Starting from Lagrangian (\ref{2}), one obtains in the leading order
of the large $N_c$-expansion (i.e. in the one-fermion loop
approximation) the following path integral expression for the
effective action ${\cal S}_{\rm {eff}}(\sigma,\pi_a)$ of the bosonic
$\sigma (x)$ and $\pi_a (x)$ fields:
$$
\exp(\mathrm{i}{\cal S}_{\rm {eff}}(\sigma,\pi_a))=
  N'\int[d\bar q][dq]\exp\Bigl(\mathrm{i}\int\tilde{\cal L}\,d^2 x\Bigr),
$$
where
\begin{equation}
{\cal S}_{\rm {eff}}
(\sigma,\pi_a)
=-N_c\int d^2x\left [\frac{\sigma^2+\pi^2_a}{4G}
\right ]+\tilde {\cal S}_{\rm {eff}},
\label{3}
\end{equation}
and $N'$ is a normalization constant.
The quark contribution to the effective action, i.e. the term
$\tilde {\cal S}_{\rm {eff}}$ in (\ref{3}), is given by:
\begin{equation}
\exp(\mathrm{i}\tilde {\cal S}_{\rm {eff}})=N'\int [d\bar
q][dq]\exp\Bigl(\mathrm{i}\int\Big\{\bar q\big
[\gamma^\rho\mathrm{i}\partial_\rho +\mu\gamma^0+
\nu\tau_3\gamma^0-\sigma -\mathrm{i}\gamma^5\pi_a\tau_a\big
]q\Big\}d^2 x\Bigr).
 \label{4}
\end{equation}
The ground state expectation values  $\vev{\sigma(x)}$ and
$\vev{\pi_a(x)}$, of the composite bosonic fields are determined by
the saddle point equations,
\begin{eqnarray}
\frac{\delta {\cal S}_{\rm {eff}}}{\delta\sigma (x)}=0,~~~~~
\frac{\delta {\cal S}_{\rm {eff}}}{\delta\pi_a (x)}=0,~~~~~
\label{5}
\end{eqnarray}
where $a=1,2,3$. In vacuum, i.e. in the state corresponding to an
empty space with zero particle density and zero values of the
chemical potentials $\mu$ and $\mu_I$, the quantities
$\vev{\sigma(x)}$ and $\vev{\pi_a(x)}$ do not depend on space
coordinates. However, in a dense medium, when $\mu\ne 0$, $\mu_I\ne
0$, the ground state expectation values of bosonic fields might have a
nontrivial dependence on $x$. In particular, in this paper we will
use the following ansatz:
\begin{eqnarray}
\vev{\sigma(x)}=M\cos (2bx),~~~\vev{\pi_3(x)}=M\sin
(2bx),~~~\vev{\pi_1(x)}=\Delta,~~~ \vev{\pi_2(x)}=0, \label{6}
\end{eqnarray}
where $M,b,$ and $\Delta$ are constant quantities. In fact, they are
coordinates of the global minimum point of the thermodynamic potential (TDP) $\Omega (M,b,\Delta)$.
\footnote{Here and in the following we will use a rather conventional notation 
"global" minimum in the sense that among all our numerically found
local minima the thermodynamical potential takes in their case the
lowest value. This does not exclude the possibility that there exist 
other inhomogeneous condensates, different from (\ref{6}), which 
lead to ground states with even lower values of the TDP.}
In the leading order of the large $N_c$-expansion it is defined by the following expression:
\begin{equation}
\int d^2x \Omega (M,b,\Delta)=-\frac{1}{N_c}{\cal S}_{\rm {eff}}\{\sigma(x),\pi_a(x)\}\big|_{\sigma
    (x)=\vev{\sigma(x)},\pi_a(x)=\vev{\pi_a(x)}} ,
\end{equation}
which gives
\begin{equation}
\mathrm{i}\int d^2x\Omega (M,b,\Delta)\,\,=\,\,\mathrm{i}\int
d^2x\frac{M^2+\Delta^2}{4G}-\frac{1}{N_c}\ln\left (
\int [d\bar q][dq]\exp\Bigl(\mathrm{i}\int d^2
x\bar q {\cal D} q \Bigr)\right )
\label{8}
\end{equation}
where
\begin{equation}
{\cal D}=\gamma^\rho\mathrm{i}\partial_\rho +\mu\gamma^0+
\nu\tau_3\gamma^0-M\exp(2\mathrm{i}\gamma^5\tau_3bx)
-\mathrm{i}\gamma^5\tau_1\Delta.\label{9}
\end{equation}
To proceed, let us introduce the new quark fields,
$q_w=\exp(\mathrm{i}\gamma^5\tau_3bx)q$ and $\bar q_w= \bar
q\exp(\mathrm{i}\gamma^5\tau_3bx)$, such that
\begin{equation}
\bar q {\cal D}q=\bar q_w \left [\gamma^\rho\mathrm{i}\partial_\rho
+\mu\gamma^0+ (b+\nu)\tau_3\gamma^0-M
-\mathrm{i}\gamma^5\tau_1\Delta\right ]q_w\equiv \bar q_w
Dq_w,\label{10}
\end{equation}
where instead of the $x-$dependent Dirac operator (\ref{9}) a new
$x-$independent operator appears
\begin{equation}
D\,=\,\gamma^\rho\mathrm{i}\partial_\rho
+\mu\gamma^0+ (b+\nu)\tau_3\gamma^0-M
-\mathrm{i}\gamma^5\tau_1\Delta.
\end{equation}
Since this transformation of quark fields does not change the path
integral measure in (\ref{8})  \footnote{This nontrivial fact
follows from the investigations by Fujikawa \cite{fujikawa}, who
established that a chiral transformation of spinor fields changes
the path integral measure only in the case when there is an
interaction between spinor and gauge fields.},  expression (\ref{8})
for the thermodynamic potential is easily transformed into the
following one:
\begin{eqnarray}
\Omega (M,b,\Delta)&=&\frac{M^2+\Delta^2}{4G}+\mathrm{i}\frac{{\rm
Tr}_{sfx}\ln D}{N_c\int d^2x}\nonumber\\
&=&\frac{M^2+\Delta^2}{4G}+\mathrm{i}{\rm
Tr}_{sf}\int\frac{d^2p}{(2\pi)^2}\ln\Big (\not\!p +\mu\gamma^0 +
(b+\nu)\tau_3\gamma^0-M-\mathrm{i}\gamma^5\Delta\tau_1\Big ),
\label{11}
\end{eqnarray}
where the Tr-operation ${\rm Tr}_{sfx}$ stands for the trace in
spinor- ($s$), flavor- ($f$) as well as two-dimensional coordinate-
($x$) spaces, respectively, and ${\rm Tr}_{sf}$ is the respective
trace without $x-$space. Since the thermodynamic potential
(\ref{11}) is formally equal to the TDP (9) of  paper \cite{ektz}
when one performs the replacement $\nu\to b+\nu$,  one can further
use the corresponding techniques and obtains
\begin{eqnarray}
\Omega(M,b,\Delta)
=\frac{M^2+\Delta^2}{4G}+\mathrm{i}\int\frac{d^2p}{(2\pi)^2}\ln
\Big\{\Big [(p_0+\mu)^2-(E^+_{\Delta})^2\Big ]\Big
[(p_0+\mu)^2-(E^-_{\Delta})^2\Big ]\Big\}, \label{12}
\end{eqnarray}
where
\begin{eqnarray}
E_\Delta^\pm=\sqrt{(E^\pm)^2+\Delta^2},~~~ E^\pm=E\pm (b+\nu),~~~
E=\sqrt{p_1^2+M^2}. \label{13}
\end{eqnarray}
The argument of the $\ln (x)$-function in (\ref{12}) is proportional
to the inverse quark propagator in the energy-momentum space representation. Hence, 
its zeros are the poles of the quark propagator. So, using (\ref{12}) one can find
the dispersion laws for quasiparticles, i.e. the momentum dependence
of the quark ($p_{0u}$, $p_{0d}$) and antiquark ($p_{0\bar u}$,
$p_{0\bar d}$) energies, in a medium (the full expression of the
quark propagator matrix is presented in Appendix B of paper \cite{ek3}):
\begin{equation}
p_{0u}=E_\Delta^--\mu,~~~p_{0d}=E_\Delta^+-\mu,~~
p_{0\bar u}=-(E_\Delta^++\mu),~~ p_{0\bar d}=-(E_\Delta^-+\mu).
\label{E6}
\end{equation}
It is clear that expression (\ref{12}) is symmetric with respect to
the transformations $\mu\to -\mu$ and $(b+\nu)\to -(b+\nu)$
respectively. Thus, without loss of generality, it is sufficient to
adopt the restrictions $\mu\ge 0$ and $(b+\nu)\ge 0$. Under these
conditions, upon integrating in (\ref{12}) over $p_0$ (see in
\cite{ektz} for similar integrals), one obtains for the TDP of the
system at zero temperature the following expression:
\begin{eqnarray}
\Omega (M,b,\Delta)&=&\frac{M^2+\Delta^2}{4G}-\int_{0}^{\infty}\frac{dp_1}{\pi}
\Big\{E_\Delta^++E_\Delta^-\nonumber\\
&+&(\mu-E_\Delta^+)\theta(\mu-E_\Delta^+)+(\mu-E_\Delta^-)\theta(\mu-E_\Delta^-)\Big\},
\label{14}
\end{eqnarray}
where $\theta(x)$ is the Heaviside theta-function.

\section{Homogeneous chiral condensate, $b=0$}
\subsection{The case with flavor symmetry,  $\mu_I=0$}

First of all let us consider the vacuum case, i.e. when
$\mu=0,\mu_I=0$ and temperature is zero. Since in QCD parity is not broken in the vacuum, it
is necessary in all QCD-motivated theories to adopt the same
requirement. In the framework of  our model this means that one
should take $\Delta=0$ if $\mu=0,\mu_I=0$. Assuming homogeneity of
the chiral condensate ($b=0$), we then obtain from (\ref{14}) the
following expression for the effective potential of the initial
NJL$_2$ model in vacuum ($\Delta=0$, $\mu=0,\mu_I=0$) \footnote{In
vacuum, the thermodynamic potential is usually called effective
potential.}:
 \begin{eqnarray}
V_{0}(M)&=&\frac{M^2}{4G}-\frac{2}{\pi}\int_{0}^{\infty}dp_1\sqrt{p_1^2+M^2}.
\label{15}
\end{eqnarray}
Formally, the effective potential (\ref{15}) is an ultraviolet (UV)
divergent quantity. To renormalize $V_0(M)$, i.e. to obtain a finite
expression for it, we first need to regularize the integral in the
right hand side of (\ref{15}) by cutting off its integration region,
$p_1<\Lambda$. Second, we suppose that the bare coupling constant
$G$ in (\ref{15}) depends on the cutoff parameter $\Lambda$
($G\equiv G(\Lambda)$) in such a way that in the limit
$\Lambda\to\infty$ one obtains a finite expression. To construct the
function $G(\Lambda)$, let us suppose that the stationarity equation
$\partial V_0(M)/\partial M=0$ has a nontrivial solution $M_0$. Then
it is easy to obtain from this equation the following expression for
the bare coupling constant $G(\Lambda)$:
\begin{eqnarray}
\frac{1}{2G(\Lambda)}=\frac{2}{\pi}\int_{0}^\Lambda
dp_1\frac{1}{\sqrt{M_0^2+p_1^2}}=\frac{2}{\pi}\ln\left
(\frac{\Lambda+\sqrt{M_0^2+\Lambda^2}}{M_0}\right ). \label{16}
\end{eqnarray}
Now, using (\ref{16}) in the regularized expression (\ref{15}) and adding an unessential
constant $\Lambda^2/\pi$, one can find at $\Lambda\to\infty$:
\begin{eqnarray}
V_{0}(M)&\equiv&\lim_{\Lambda\to\infty}\left\{\frac{M^2}{4G(\Lambda)}-
\frac{2}{\pi}\int_{0}^{\Lambda}dp_1\sqrt{p_1^2+M^2}+\frac{\Lambda^2}{\pi}\right\}=
\frac{M^2}{2\pi}\left [\ln\left
(\frac{M^2}{M_0^2}\right )-1\right ]. \label{17}
\end{eqnarray}
Since $M_0$ might be considered as a free model parameter, it
follows from (\ref{16}) and (\ref{17}) that the renormalization
procedure of the NJL$_2$ model is accompanied by the dimensional
transmutation phenomenon. Indeed, in the initial unrenormalized
expression (\ref{15}) for $V_0(M)$ the dimensionless coupling
constant $G$ is present, whereas after renormalization the effective
potential (\ref{17}) is characterized by a dimensional free model
parameter $M_0$. Moreover, as it is clear from (\ref{17}), the
global minimum point of the effective potential $V_0(M)$ lies just
at the point $M=M_0$, so in vacuum the chiral $SU_L(2)\times
SU_R(2)$ symmetry of the NJL$_2$ model (1) is always spontaneously
broken and the quantity $M_0$ might be treated as dynamical quark
mass (in vacuum).

A detailed information about the phase structure of the NJL$_2$
model (1) at $\mu\not = 0,\mu_I=0$ can be found, e.g., in
\cite{wolff}. So, at $\mu>M_0/\sqrt{2}$ and $\mu_I=0$ there is a
massless chirally symmetric phase with nonzero baryon density.
However, at $\mu<M_0/\sqrt{2}$ and $\mu_I=0$ chiral symmetry is
spontaneously broken down and quarks acquire a mass $M_0$. In this
phase baryon density is equal to zero.

\subsection{Phase structure in the general case: $\mu\ne 0,\mu_I\ne 0$}

To find the phase portrait of the NJL$_2$ model (1) in the case of a
homogeneous chiral condensate but for arbitrary values of chemical
potentials and at zero temperature, one should start from the expression (\ref{14}) with
$b=0$. (Note, that at $\mu_I\ne 0$ the condensation of charged pions
might occur, so we need to take into account a nonzero value of
$\Delta$.) Obviously, this expression is again UV-divergent, so
first of all it is necessary to regularize it. Using, as the most
simple regularization, a $\Lambda$-cutoff in the one-dimensional
momentum space, we have:
\begin{eqnarray}
\Omega_{\rm{reg}}(M,b=0,\Delta)&=&\frac{M^2+\Delta^2}{4G}-\int_{0}^{\Lambda}
\frac{dp_1}{\pi}\Big\{{\cal E}_\Delta^++{\cal E}_\Delta^-\Big\}\nonumber\\
&-&\int_{0}^{\infty}\frac{dp_1}{\pi}\Big\{(\mu-{\cal E}_\Delta^+)\theta(\mu-{\cal E}_\Delta^+)+
(\mu-{\cal E}_\Delta^-)\theta(\mu-{\cal E}_\Delta^-)\Big\},
\label{18}
\end{eqnarray}
where ${\cal E}_\Delta^\pm$ denotes the quantity $E_\Delta^\pm$
(\ref{13}) at $b=0$. Due to the presence of $\theta$-functions, the
second integral in (\ref{18}) has a finite integration region, i.e.
it is a proper integral not needed to be regularized. To obtain a
finite (renormalized) expression $\Omega (M,\Delta)$ for the
thermodynamic potential, one should again perform in (\ref{18}) the
replacement $G\to G(\Lambda)$, the last quantity being given in
(\ref{16}),  and then let $\Lambda$ tend  to infinity (compare with
(\ref{17})), i.e.
\begin{eqnarray}
\Omega (M,\Delta)=\lim_{\Lambda\to\infty}\left\{\Omega_{\rm{reg}}
(M,b=0,\Delta)\Big |_{G\to G(\Lambda)}+\frac{\Lambda^2}{\pi}\right\}.
\label{19}
\end{eqnarray}
Using the definition of the effective potential in vacuum (see
(\ref{17})), it is easy to obtain the following renormalization
invariant expression of the TDP (\ref{19})
\begin{eqnarray}
\Omega (M,\Delta)&&=V_0(\sqrt{M^2+\Delta^2})-\int_{0}^{\infty}
\frac{dp_1}{\pi}\Big\{{\cal E}^+_{\Delta}+{\cal E}^-_{\Delta}-2\sqrt{p_1^2+M^2+
\Delta^2}\Big\}\nonumber\\
&&-\int_{0}^{\infty}\frac{dp_1}{\pi}\Big\{(\mu-{\cal E}^+_{\Delta})
\theta(\mu-{\cal E}^+_{\Delta})+(\mu-{\cal E}^-_{\Delta})\theta
(\mu-{\cal E}^-_{\Delta})\Big\}, \label{20}
\end{eqnarray}
where the function $V_0(x)$ is defined in (\ref{17}). Moreover, the
second integral in (\ref{20}) is proper (see also the corresponding
remark just after (\ref{18})), whereas the first integral is
convergent and defined as
\begin{eqnarray}
\int_{0}^{\infty}
dp_1\Big [{\cal E}^+_{\Delta}+{\cal E}^-_{\Delta}-2\sqrt{p_1^2+M^2+
\Delta^2}~\Big ]
= \lim_{\Lambda\to\infty}\left\{\int_{0}^{\Lambda}
dp_1\Big [{\cal E}^+_{\Delta}+{\cal E}^-_{\Delta}-2\sqrt{p_1^2+M^2+
\Delta^2}~\Big ]\right\}.\label{21}
\end{eqnarray}
Thus, in the case of a homogeneous chiral condensate, the TDP is
given by (\ref{19})-(\ref{20}) and the corresponding phase
structure, following from it, is  depicted in Fig. 1 (for a more
detailed investigation of this TDP see paper \cite{ek2}).
There, the phases denoted by 1 and 2 correspond to the global
minimum point (GMP) of the form $(M=0,\Delta=0)$ and $(M\ne
0,\Delta=0)$, correspondingly. In the pion condensed phase (PC) the
GMP of the TDP (\ref{20}) has the form $(M=0,\Delta=M_0)$, i.e. in
this phase the isospin symmetry $U_{I_3}(1)$ is broken spontaneously
\footnote{Note that our numerical investigations show that the TDP
(\ref{20}) has no local minima of the form $(M\ne 0,\Delta\ne 0)$,
i.e. simultaneous dynamical quark mass generation and charged 
pion condensation are incompatible in the framework of the NJL$_2$ model (1) at $b=0$. The
same is valid for the simple two-flavored NJL$_4$ model in the
mean-field approximation \cite{ek}. However, it is not excluded that
there might be realized a mixed phase with both nonzero gaps, $M\ne 0$ and
$\Delta\ne 0$, in models with a more complicated four-fermion structure.}.
It is easy to see that throughout the PC phase the quark number density is equal to zero, whereas the isospin
density $n_I=-\partial\Omega/\partial\mu_I$ is equal to $\nu/\pi$.
\begin{figure}
 \includegraphics[width=0.45\textwidth]{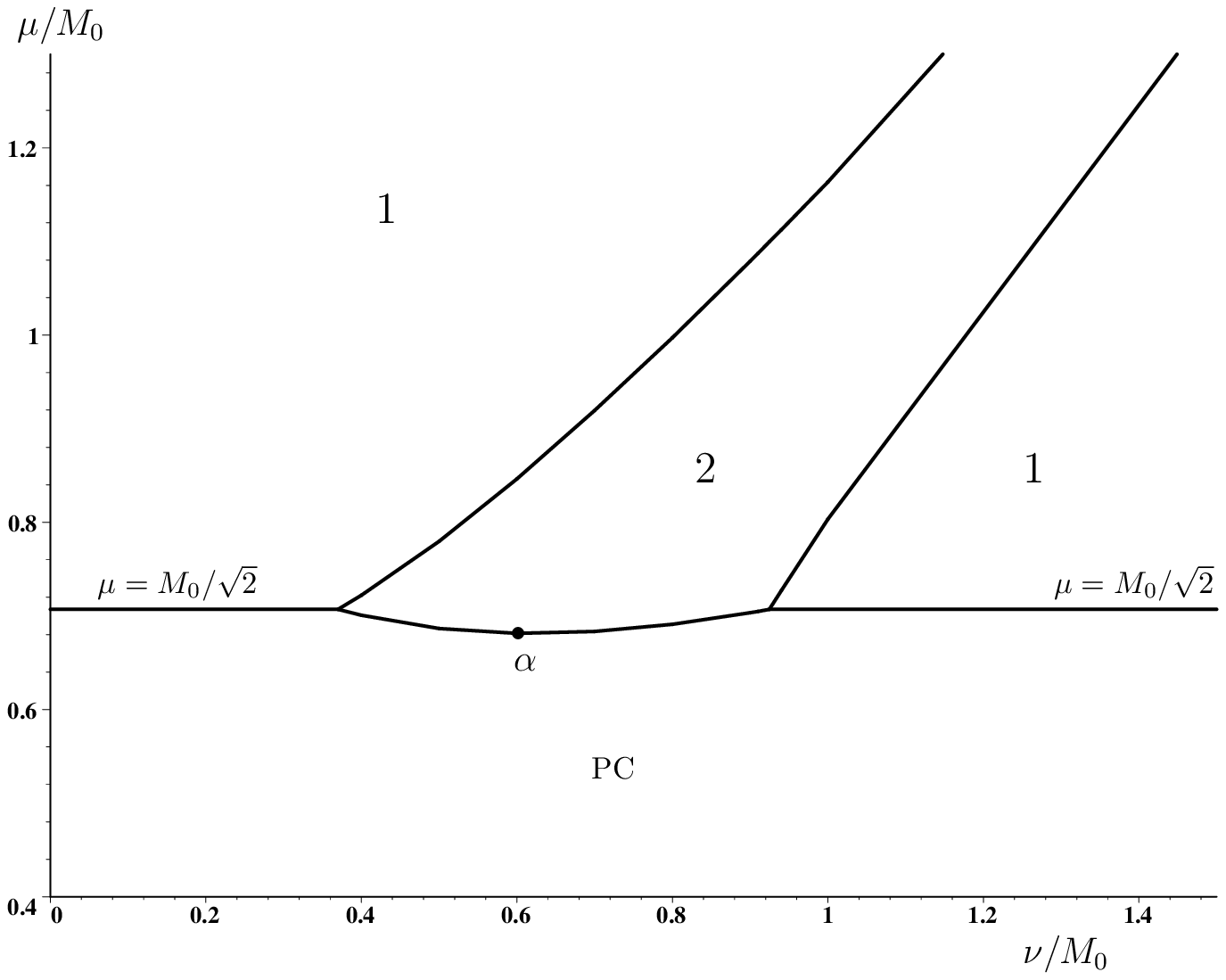}
 \hfill
\includegraphics[width=0.45\textwidth]{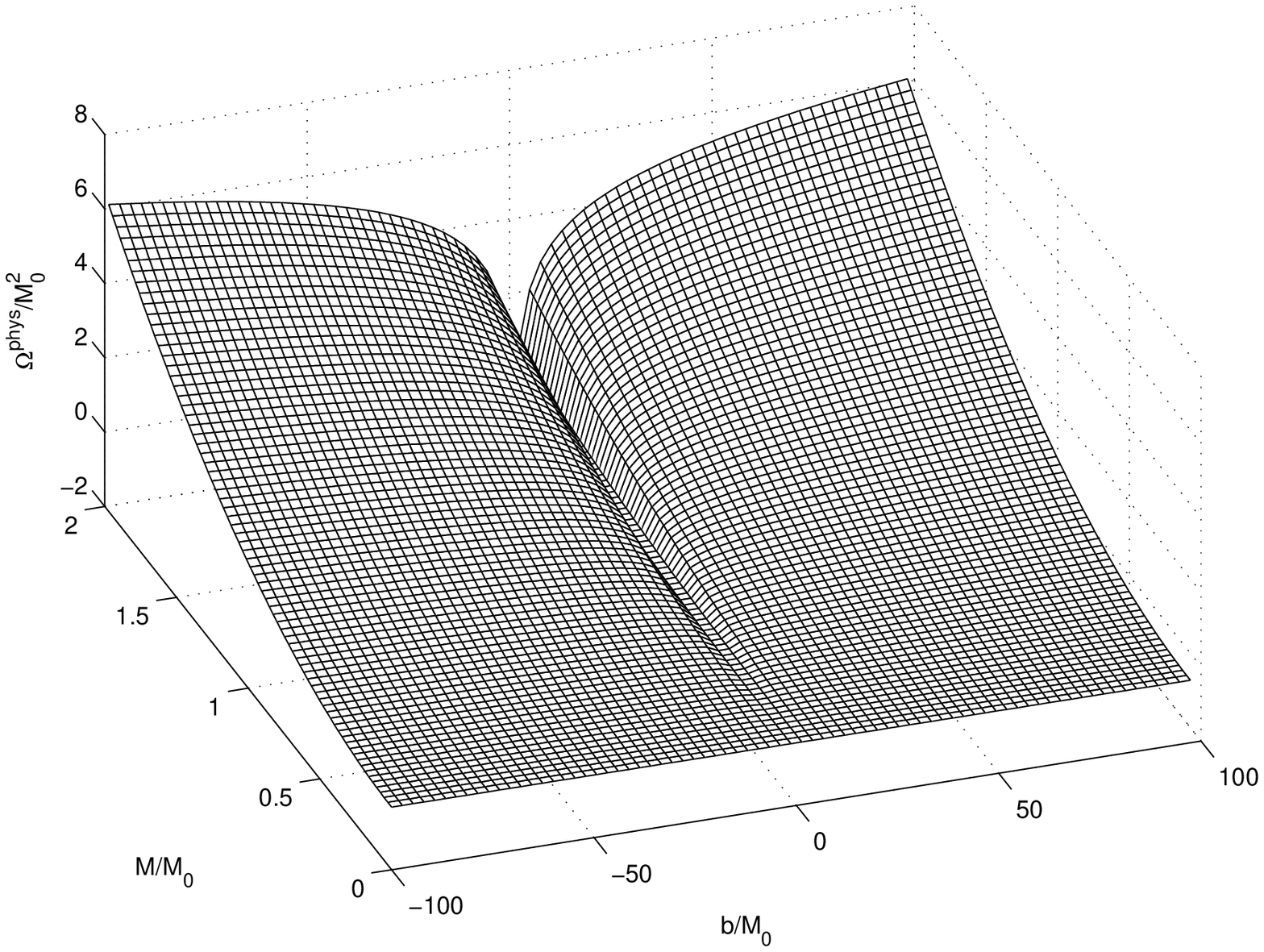}\\
\parbox[t]{0.45\textwidth}{
 \caption{The $(\mu,\nu)$ phase portrait of the model considered at
 $T=0$ and $\nu>0$ in the case of spatially homogeneous condensates.
Here  $\nu=\frac{\mu_I}{2}$, and  $M_0$ is the quark mass in the vacuum.
Number 1 denotes the  symmetric phase with massless quarks, number 2  the normal 
quark  matter phase with massive quarks, and PC denotes the charged pion condensed 
phase. The point $\alpha$  is the lowest point of the phase 2: $\mu_\alpha\approx 0.68M_0$, $\nu_\alpha\approx 0.6M_0$.} }
\hfill
\parbox[t]{0.45\textwidth}{
\caption{The plot of $\Omega^{\rm{phys}}$ (\ref{25}) vs $M,b$ at
$\mu_I=0,\mu=M_0,\Delta=0$. } }
\end{figure}

\section{Inhomogeneous chiral condensate, $b\ne 0$}

To obtain the phase portrait of the initial NJL$_2$ model in this
case (temperature $T$ is zero), let us start from the most general expression for the TDP
(\ref{14}). As previously, let us first use  the most simple
momentum cutoff regularization of this quantity,
\begin{eqnarray}
\Omega_{\rm{reg}}(M,b,\Delta)&=&\frac{M^2+\Delta^2}{4G}-\int_{0}^{\Lambda}
\frac{dp_1}{\pi}\Big\{E_\Delta^++E_\Delta^-\Big\}\nonumber\\
&-&\int_{0}^{\infty}\frac{dp_1}{\pi}\Big\{(\mu-E_\Delta^+)\theta(\mu-E_\Delta^+)+
(\mu-E_\Delta^-)\theta(\mu-E_\Delta^-)\Big\},
\label{22}
\end{eqnarray}
where the expressions for $E_\Delta^\pm$ are presented in
(\ref{13}). The corresponding renormalized expression for the TDP is
again defined by (compare with (\ref{19}))
\begin{eqnarray}
\Omega (M,b,\Delta)=\lim_{\Lambda\to\infty}\left\{\Omega_{\rm{reg}}
(M,b,\Delta)\Big |_{G\to G(\Lambda)}+\frac{\Lambda^2}{\pi}\right\},
\label{23}
\end{eqnarray}
where $G(\Lambda)$ is given in (\ref{16}), and reads
\begin{eqnarray}
\Omega (M,b,\Delta)&&=V_0(\sqrt{M^2+\Delta^2})-\lim_{\Lambda\to\infty}\left\{
\int_{0}^{\Lambda}
\frac{dp_1}{\pi}\left [E^+_{\Delta}+E^-_{\Delta}-2\sqrt{p_1^2+M^2+
\Delta^2}\right ]\right\}\nonumber\\
&&-\int_{0}^{\infty}\frac{dp_1}{\pi}\Big\{(\mu-E^+_{\Delta})
\theta(\mu-E^+_{\Delta})+(\mu-E^-_{\Delta})\theta
(\mu-E^-_{\Delta})\Big\}. \label{230}
\end{eqnarray}
(Evidently, at $b=0$ this expression coincides with the TDP $\Omega(M,\Delta)$ (\ref{20}).)
The global minimum point of the function $\Omega (M,b,\Delta)$
(\ref{230}) vs variables $M,b$ and $\Delta$ should render the phase
structure of the model. However, two circumstances prevent us from
considering this quantity as a genuine physical thermodynamic
potential of the system. The first is that the function (\ref{230})
is not bounded from below with respect to the variable $b$. Secondly, it
is intuitively clear  that at $M=0$ the genuine thermodynamic
potential should not depend on the variable $b$, because no
observable quantity may depend on a wave vector if the amplitude of
the corresponding oscillations (wave) is zero. However, the TDP
defined by (\ref{230}) at $M=0$ (see also in \cite{ek2})
\begin{eqnarray}
\Omega(M=0,b,\Delta)=
V_0(\Delta)-\frac{(b+\nu)^2}{\pi}+\frac{\theta(\mu-\Delta)}{\pi}
\left [\Delta^2\ln\left
(\frac{\mu+\sqrt{\mu^2-\Delta^2}}{\Delta}\right
)-\mu\sqrt{\mu^2-\Delta^2}\right ], \label{24}
\end{eqnarray}
retains an unphysical  dependence on $b$. Clearly, the
two above mentioned unphysical properties of the TDP (\ref{230}) are
due to the term $-\frac{(b+\nu)^2}{\pi}$ in (\ref{24}). Hence, the
subtraction of this term from the TDP (\ref{230}) brings us to the
quantity, which might  serve as a physically acceptable
thermodynamic potential of the system,
\begin{eqnarray}
\Omega^{\rm{phys}} (M,b,\Delta)=\Omega
(M,b,\Delta)+\frac{(b+\nu)^2}{\pi}-\frac{\nu^2}{\pi}. \label{25}
\end{eqnarray}
(We also add in the expression (\ref{25}) a $b$-independent term, -$\nu^2/\pi$, in order to
reproduce at $b=0$ the TDP (\ref{20}), corresponding to a spatially homogeneous chiral condensate.)
The reason why the expression for the TDP (\ref{25}) does not follow
straightforwardly from the unrenormalized TDP expression (\ref{14})
lies in the usage of the {\it symmetric momentum cutoff regularized}
TDP (\ref{22}). This means that for each  
energy $E_\Delta^\pm$ the integration in the first (regularized) integral
of (\ref{22}) is performed over the same momentum interval
$0<p_1<\Lambda$. Correspondingly, in this case there is an asymmetry
in values of energies $E_\Delta^\pm$, which contribute
to $\Omega_{\rm{reg}}(M,b,\Delta)$. Indeed, if $p_1<\Lambda$, then
$E_\Delta^\pm <\sqrt{\left (\sqrt{\Lambda^2+M^2}\pm (b+\nu)\right
)^2+\Delta^2}$, i.e. for different quasiparticles there are allowed
different regions of their energy values. However, as discussed in
the recent papers \cite{nakano,nickel} a more adequate
regularization scheme in the case of spatially inhomogeneous phases
is that one, where there is an energy constraint which is the same
for all quasiparticles. So, dealing with spatial inhomogeneity, one
can use, e.g., the Schwinger proper-time regularization, dimensional
regularization etc. In particular, in the recent paper \cite{zfk}
the {\it symmetric energy cutoff regularization} scheme was proposed
in considering the behavior of chiral density waves in the presence
of an external magnetic field in the framework of a four-dimensional
Nambu--Jona-Lasinio model. There, for each quasiparticle the same
(finite) interval of their energy values was allowed to contribute
to the regularized thermodynamic potential. As a result, a
physically relevant renormalized TDP  without the above-mentioned
shortcomings was obtained.

In this paper the slightly modified energy cutoff regularization
scheme of \cite{zfk} is adopted. Namely, we require that only
energies with momenta $p_1$, constrained by the
relations  $E_\Delta^\pm(M=0,\Delta=0)=p_1\pm(b+\nu)<\Lambda$,
contribute to the regularized thermodynamic potential. This means
that the term with energy $E_\Delta^+$ ($E_\Delta^-$)
should be integrated in the regularized expression for TDP over the
interval $0<p_1<\Lambda-(b+\nu)$ ($0<p_1<\Lambda+(b+\nu)$). (A
similar regularization was used in studying the CDW phase in a
two-dimensional NJL model without isospin chemical potential
\cite{ohwa}.) Consequently, we have the following regularized
expression for the TDP (\ref{14})
\begin{eqnarray}
\widetilde\Omega_{\rm{reg}}(M,b,\Delta)&=&\frac{M^2+\Delta^2}{4G}-\frac
1\pi\int_{0}^{\Lambda-\tilde\nu} dp_1~E_\Delta^+-\frac
1\pi\int_{0}^{\Lambda+\tilde\nu}
dp_1~E_\Delta^-\nonumber\\
&-&\int_{0}^{\infty}\frac{dp_1}{\pi}\Big\{(\mu-E_\Delta^+)\theta(\mu-E_\Delta^+)+
(\mu-E_\Delta^-)\theta(\mu-E_\Delta^-)\Big\},
\label{26}
\end{eqnarray}
where $\tilde\nu=(b+\nu)$. Replacing in this formula $G$ by
$G(\Lambda)$ from (\ref{16}) and adding an unessential constant
$(\Lambda^2-\nu^2)/\pi$, we obtain a physically ``improved'' renormalized expression
$\widetilde\Omega(M,b,\Delta)$ for the TDP (\ref{14}) when
$\Lambda\to \infty$, which differs from the expression
$\Omega(M,b,\Delta)$ in (\ref{230}). Comparing (\ref{22}) and
(\ref{26}) one can easily find that
\begin{eqnarray}
\widetilde\Omega(M,b,\Delta)-\Omega(M,b,\Delta)=\lim_{\Lambda\to\infty}\left\{
\frac 1\pi\int_{\Lambda-\tilde\nu}^\Lambda dp_1~E_\Delta^+-\frac
1\pi\int_{\Lambda}^{\Lambda+\tilde\nu}
dp_1~E_\Delta^-\right\}=\frac{(b+\nu)^2}{\pi}-\frac{\nu^2}{\pi}. \label{27}
\end{eqnarray}
(To obtain the last expression in (\ref{27}) one should take into
account that at $\Lambda\to\infty$ the $p_1$-values in both
integrals are much greater than $M,\Delta,b,\mu,\mu_I$. In this case
it is possible to expand the quantities $E_\Delta^\pm$ into power
series of $p_1$ and then to integrate each term.) Comparing
(\ref{25}) and (\ref{27}), we see that
$\widetilde\Omega(M,b,\Delta)=\Omega^{\rm{phys}} (M,b,\Delta)$, i.e.
there exists a regularization scheme \footnote{Moreover, we expect
that any regularization scheme, in which there is a constraint on
the energy values common for all quasiparticles, should provide us
with TDP $\Omega^{\rm{phys}} (M,b,\Delta)$ (\ref{25}). Among these
regularizations are dimensional and analytical ones, Pauli-Villars
and Schwinger prope-time regularizations, as well as the above
mentioned symmetric energy cutoff regularization \cite{zfk}. In
particular, the proper-time regularization is often used in studying
inhomogeneous phases in the framework of NJL models
\cite{nakano,nickel} and does not lead to any unphysical effects,
etc.}, which in the case of the inhomogeneous chiral condensate
(\ref{6}) brings us to a physically acceptable TDP
$\Omega^{\rm{phys}} (M,b,\Delta)$ (\ref{25}). Notice also that if
$b=0$ then $\Omega(M,b,\Delta)$ is equal to $\Omega^{\rm{phys}} (M,b,\Delta)$.
Hence, in the case of homogeneous chiral condensates the two above considered
regularization schemes are equivalent. In contrast, in the
inhomogeneous case the thermodynamic potentials $\Omega(M,b,\Delta)$
and $\Omega^{\rm{phys}} (M,b,\Delta)$ differ by terms, containing
the dynamical quantity $b$. As a result, the regularizations are not
equivalent. However, since the symmetric momentum cutoff
regularization (SMCR) is easier to handle, it is possible to perform
all calculations in the framework of the SMCR scheme and then simply
correct the obtained TDP $\Omega(M,b,\Delta)$ by the terms
$\frac{(b+\nu)^2}{\pi}-\frac{\nu^2}{\pi}$ [see (\ref{25})], instead of using from the
beginning one of the physically acceptable regularizations bringing
us directly to the TDP  $\Omega^{\rm{phys}} (M,b,\Delta)$.

To illustrate the fact that the TDP $\Omega^{\rm{phys}}
(M,b,\Delta)$ is bounded from below as a function of the variable
$b$, we plot in Fig. 2 this thermodynamic potential vs $M,b$ in the
particular case $\mu_I=0,\Delta=0,\mu=M_0$.
\begin{figure}
 \includegraphics[width=0.45\textwidth]{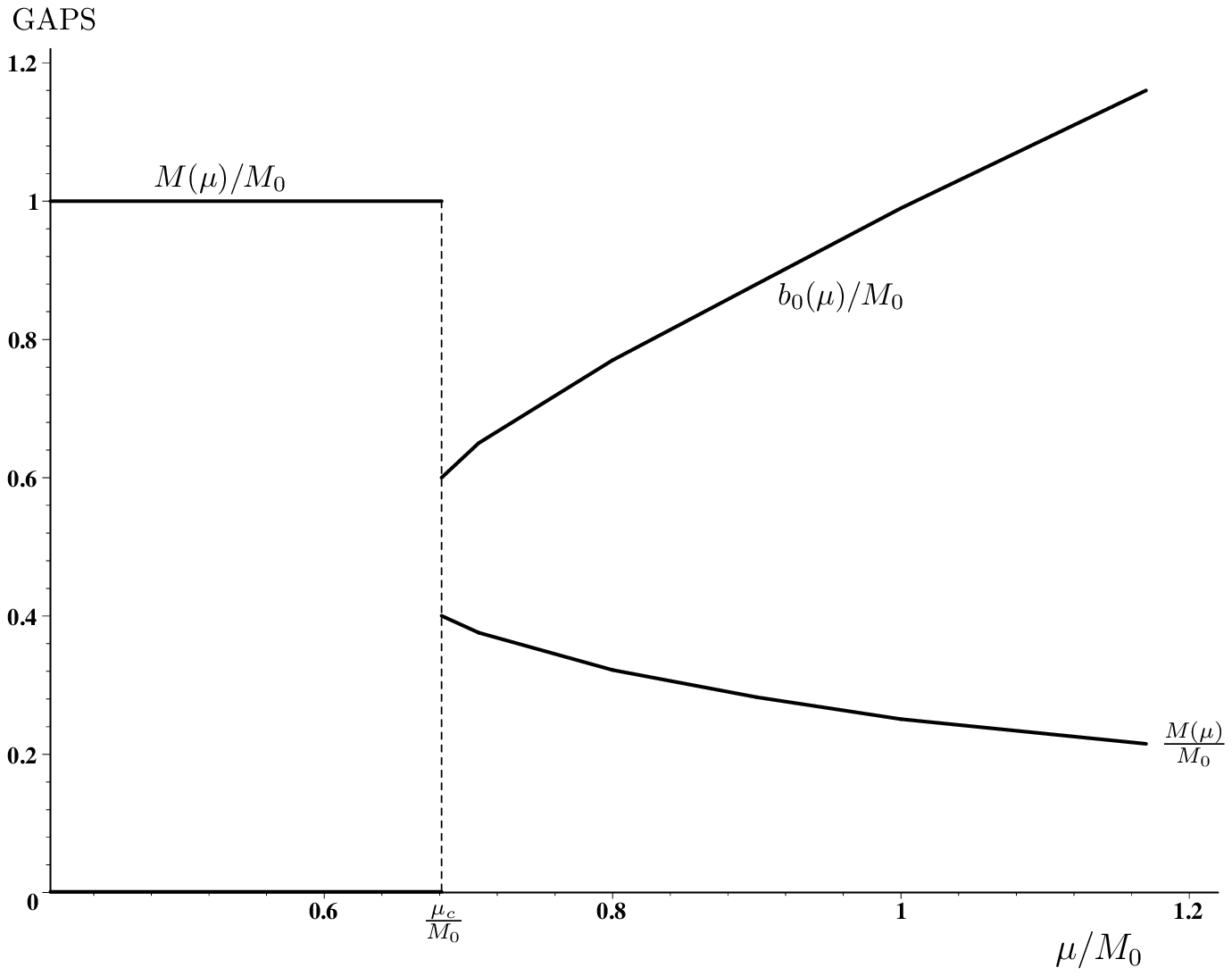}
 \hfill
\includegraphics[width=0.45\textwidth]{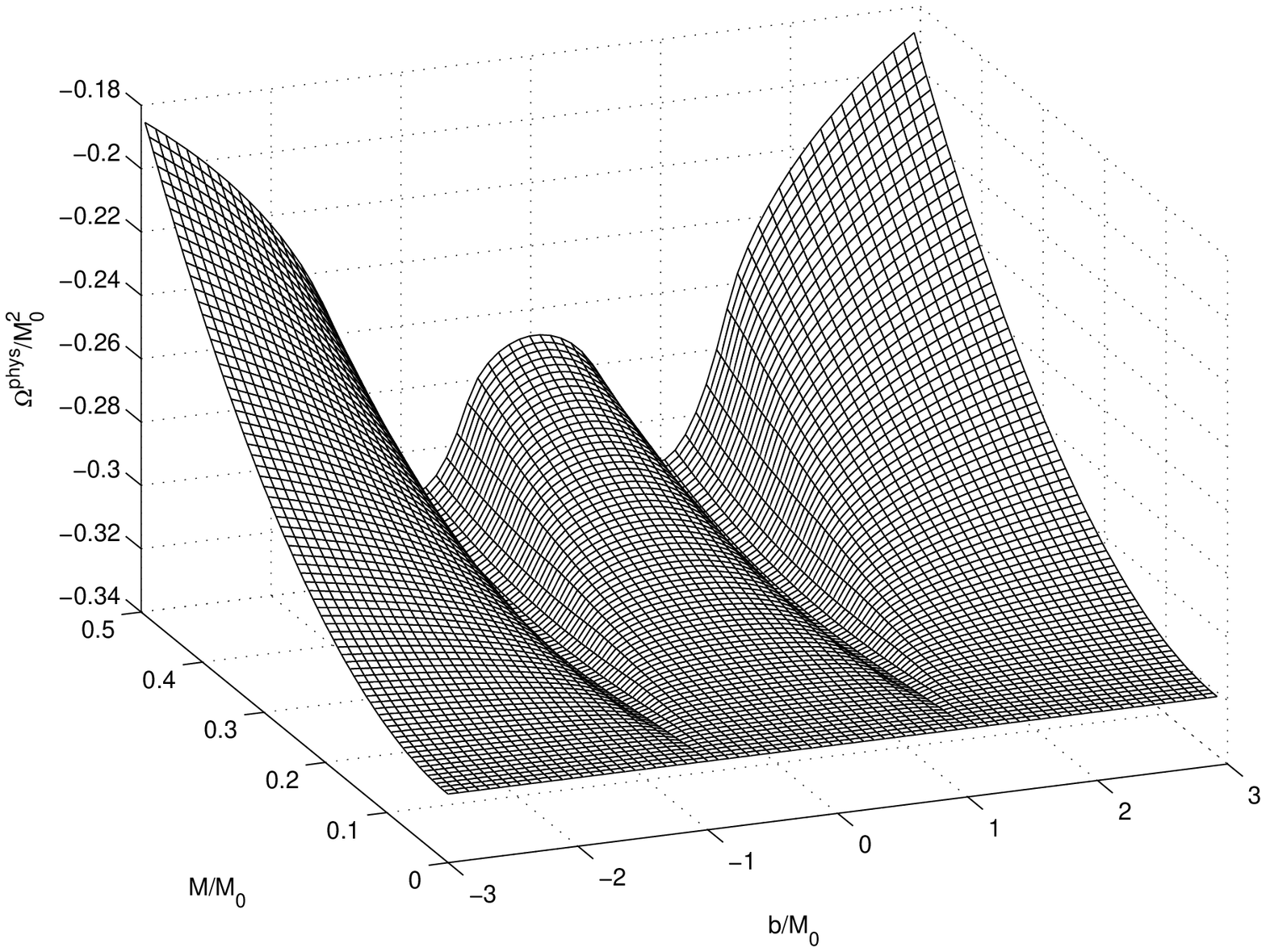}\\
\parbox[t]{0.45\textwidth}{ \caption{The CDW amplitude $M(\mu)$
and its wave vector $b_0(\mu)$ as functions of
$\mu$ in the case of zero isospin chemical potential. Here
$\mu_c=\mu_\alpha\approx 0.68M_0$.} }
\hfill
\parbox[t]{0.45\textwidth}{
\caption{The plot of $\Omega^{\rm{phys}}$ (33) vs $M,b$ at
$\mu=M_0$.}}
\end{figure}

\begin{figure}
\includegraphics[width=0.45\textwidth]{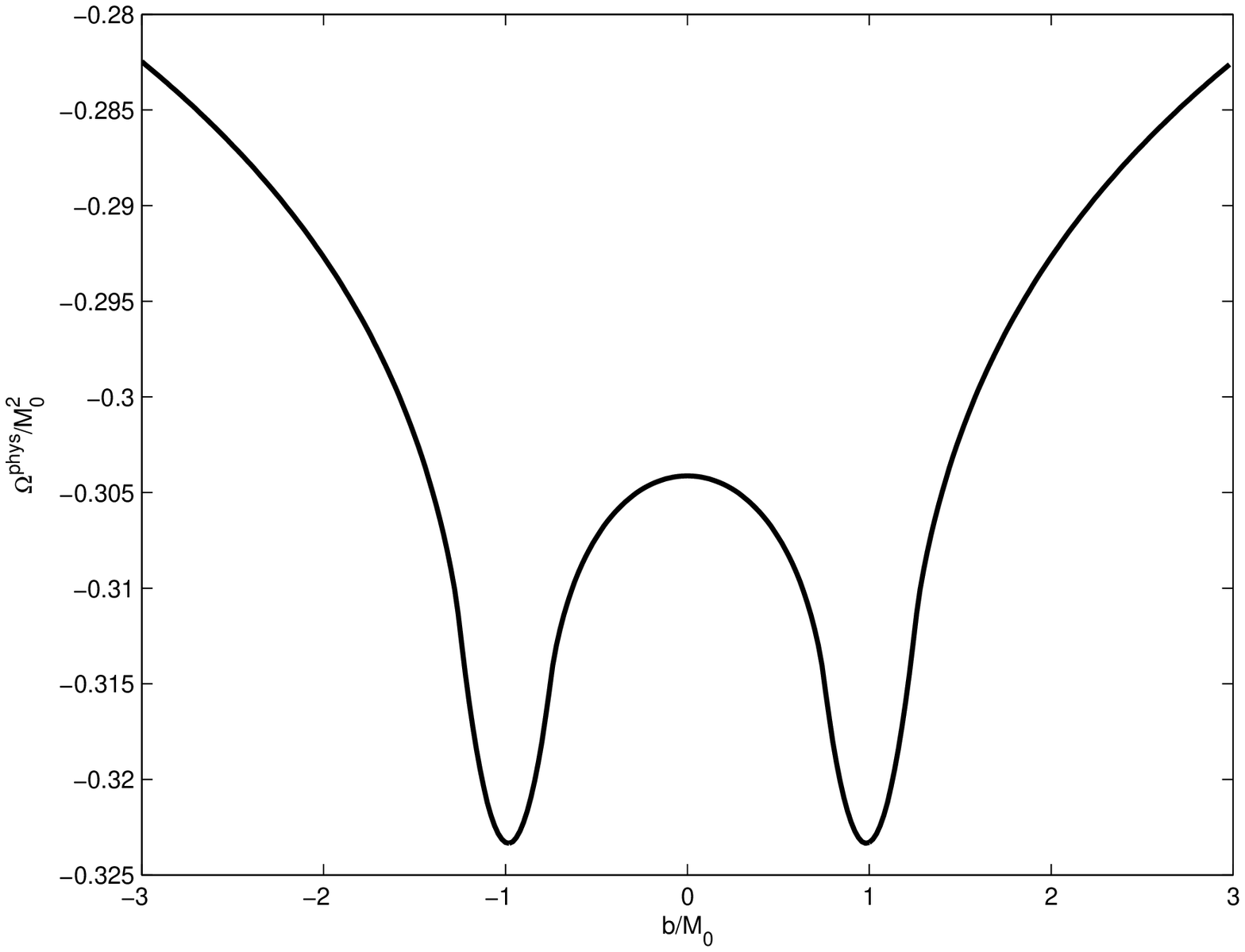}
\hfill
\includegraphics[width=0.45\textwidth]{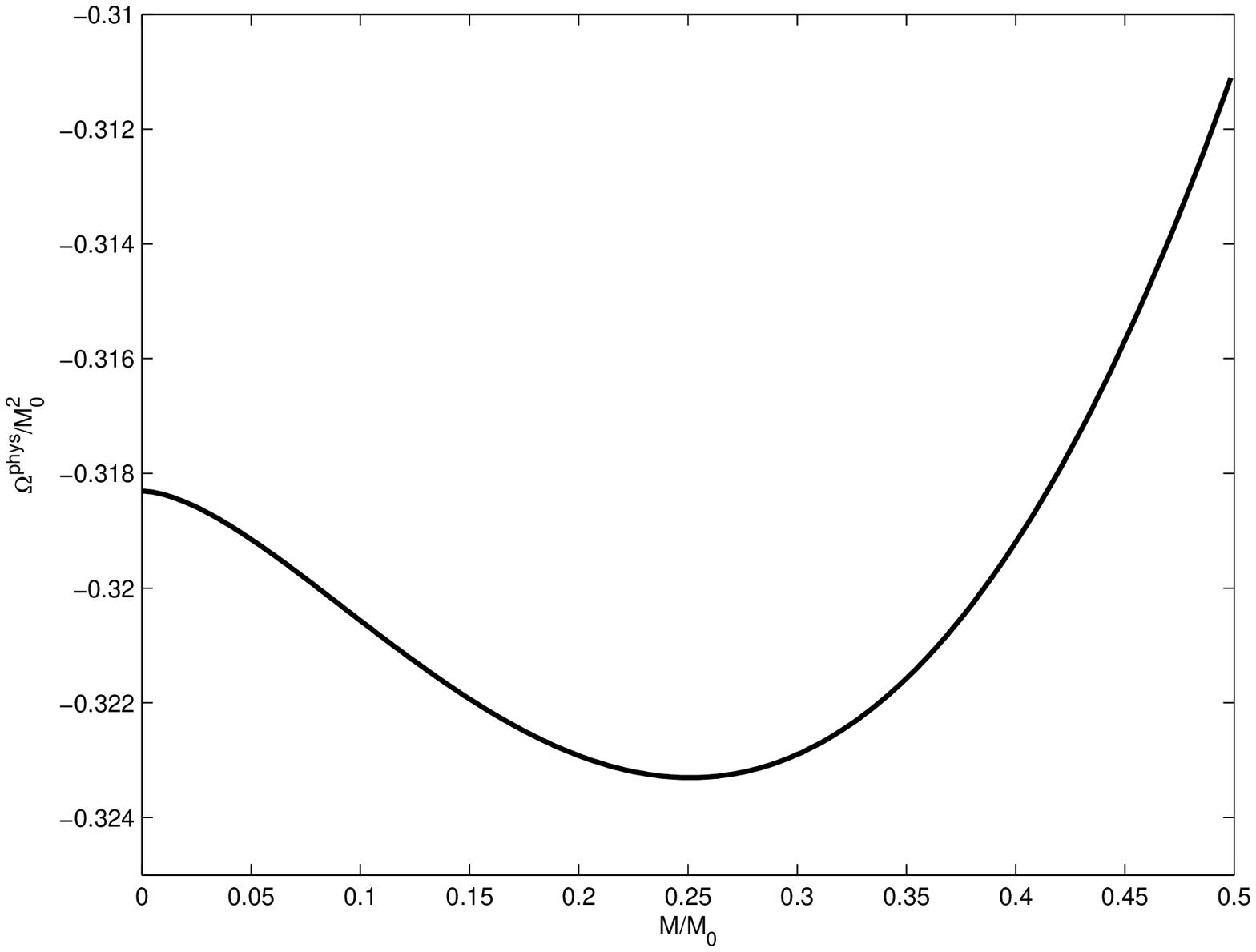}
\\
\parbox[t]{0.45\textwidth}{
 \caption{Section of the plot of $\Omega^{\rm{phys}}$ (33) vs $b$ at
$\mu=M_0$ along the axis $b$ passing through the point of minimum.} }
\hfill
\parbox[t]{0.45\textwidth}{
 \caption{Section of the plot of $\Omega^{\rm{phys}}$ (33) vs $M$ at
$\mu=M_0$ along the axis $M$ passing through the point of minimum.} }
\end{figure}

\subsection{Particular case: $\mu_I =0$, $\mu\ne 0$ }

Recall that the CDW inhomogeneous phase was established earlier in
the NJL$_2$ model with $U_L(1)\times U_R(1)$ chiral symmetry for all
$\mu>0$ at rather low temperatures \cite{ohwa,thies}. In contrast,
in this paper we are going to study chiral density waves in the
NJL$_2$ model with a continuous chiral $SU_L(2)\times SU_R(2)$
symmetry. In the present section we consider the case of $T=0$. It
is well known that at $\mu_I=0$ the charged pion condensation
phenomenon is forbidden (see, e.g., in \cite{ek2}), so without loss
of generality one may suppose that $\Delta=0$ in (\ref{230}). Then
the TDP $\Omega(M,b,\Delta=0)$ can be easily evaluated analytically
(see \cite{ek2}) and the physical thermodynamic potential
$\Omega^{\rm{phys}} (M,b)\equiv \Omega^{\rm{phys}} (M,b,\Delta=0)$
(\ref{25}) looks like
\begin{eqnarray}
\Omega^{\rm{phys}} (M,b)\!\!&=&\!\!
V_0(M)+\frac{\theta(\mu+b-M)}{2\pi} \left [M^2\ln\left
(\frac{\mu+b+\sqrt{(\mu+b)^2-M^2}}{M}\right
)-(\mu+b)\sqrt{(\mu+b)^2-M^2}\right ]\nonumber\\
&+&\frac{\theta(|\mu-b|-M)}{2\pi} \left [M^2\ln\left
(\frac{|\mu-b|+\sqrt{(\mu-b)^2-M^2}}{M}\right
)-|\mu-b|\sqrt{(\mu-b)^2-M^2} \right ]+\frac{b^2}{\pi}. \label{28}
\end{eqnarray}
Recall that in (\ref{28}) the constraints $\mu\ge 0,b\ge 0,M\ge 0$ are
supposed. The phase structure of the model in this particular case
is defined by the properties of the global minimum point (GMP) of
the TDP (\ref{28}), which certainly depend on the values of $\mu$.
The stationarity (gap) equations of this TDP, i.e. the equations
$\partial\Omega^{\rm{phys}}(M,b)/
\partial M=0$ and $\partial\Omega^{\rm{phys}}(M,b)/\partial b=0$, read:
\begin{eqnarray}
&&M\left\{\ln\left (\frac{M^2}{M_0^2}\right )+\theta
(\mu+b-M)\ln\left (\frac{\mu+b+\sqrt{(\mu+b)^2-
M^2}}{M}\right )\right.\nonumber\\
&&\left.~~~~~~~~~~~~~~~~~~~~~~~~~~~ +\theta(|\mu-b|-M)\ln\left
(\frac{|\mu-b|+\sqrt{(\mu-b)^2-M^2}}{M}\right
)\right\}=0,\label{29}\\
&&2b=\theta (\mu+b-M)\sqrt{(\mu+b)^2-M^2}+\mbox{sign}(b-\mu)
\theta(|b-\mu|-M)\sqrt{(b-\mu)^2-M^2}. \label{30}
\end{eqnarray}
Numerical investigations of the TDP (\ref{28}) and of the gap
equations (\ref{29}), (\ref{30}) show that in the NJL$_2$ model with
chiral $SU_L(2)\times SU_R(2)$ symmetry the inhomogeneous CDW phase
is realized only at $\mu>\mu_c\approx 0.68~M_0$. In contrast, at
$T=0$, in the (1+1)-dimensional $U_L(1)\times U_R(1)$ chirally
symmetric model the CDW phase appears at arbitrary nonzero values of
$\mu$ \cite{ohwa,thies}. Note that the critical value $\mu_c$ is
equal to $\mu_\alpha$ which corresponds to the lowest point of the
homogeneous phase 2 (see Fig. 1).
 Below the critical chemical potential, i.e. at $\mu<\mu_c$,
the usual homogeneous phase is arranged, where chiral symmetry is
broken down to the diagonal $SU(2)$ subgroup. The behavior of the
chiral density wave amplitude $M(\mu)$ and its wave vector
$b_0(\mu)$, which are the coordinates of the global minimum point of
the TDP (\ref{28}), is shown in Fig. 3 for $\mu_I=0$. It follows
from this figure that at the critical point $\mu_c$ a first order
phase transition takes place, since here the order parameter $M$
changes its value by a jump.
Since in the CDW phase the relation $M(\mu)<b_0(\mu)<\mu$ is valid,
it is clear from the dispersion laws (\ref{E6}) at $\Delta=0$ that
$u$-quarks are gapless excitations of this phase. It means that for
each $\mu>\mu_c$ there exist a momentum $p_1(\mu)$ at which the
quasiparticle energy $p_{0u}$ is equal to zero, i.e. there is no energy 
cost to create $u$-quarks in the system. In contrast, for the
energy of $d$-quarks we have throughout the CDW phase the relation
$p_{0d}>p_{0min}=M(\mu)+b_0(\mu)-\mu\approx M(\mu)$, i.e. there is a
gap in the energy spectrum of $d$-quarks which are called, due to
this reason, gapped excitations of the CDW phase.
There is one more peculiarity of the CDW phase. Indeed, as it is easily
seen from (\ref{28}), at $\nu=0$ the effective quark number chemical
potential of $u$-quarks is equal to $\mu +b$, whereas for $d$-quarks
it is $\mu-b$. Hence, there is a splitting of Fermi surfaces of
up/down quarks by $2b_0(\mu)$ in the CDW phase even at zero $\mu_I$.

The fact that at $\mu=M_0$ the TDP (\ref{28}) has a nontrivial
minimum at the point $(M\approx 0.25M_0,b\approx 0.99M_0)$ is well
supported by figures Fig. 4, Fig. 5 and Fig. 6, where the plot of
the function $\Omega^{\rm{phys}} (M,b)$ of $M$ and $b$ and its
sections along axes $b$ and $M$ are presented (note that in order to
draw the figures we continue the function (\ref{28}) symmetrically
onto the negative semi-axis $b$).

The influence of nonzero temperature on the formation of CDWs in the
case $\mu_I=0$ is considered, in particular, in the next section V
(see Fig. 8).

\begin{figure}
\includegraphics[width=0.45\textwidth]{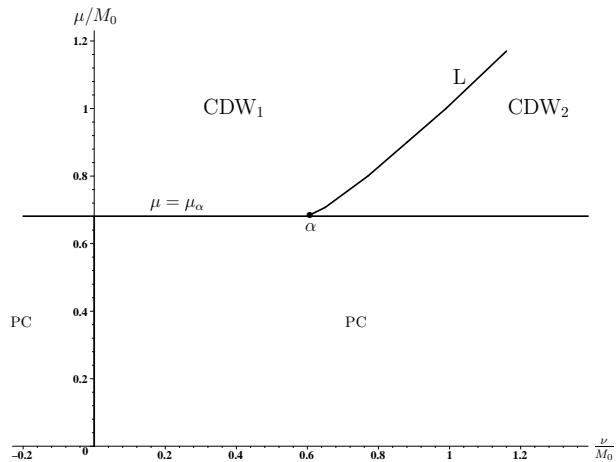}
\\
\parbox[t]{0.45\textwidth}{
\caption{The $(\nu,\mu)$ phase portrait of the model at $T=0$ when a
spatial CDW inhomogeneity is taken into account. In the CDW$_1$
(CDW$_2$) phase $b>0$ ($b<0$). The curve L, on which $b=0$,
corresponds to the homogeneous chiral symmetry  broken phase. The
same is true for
the interval $0<\mu<\mu_\alpha$ of the $\mu$-axis, where $\mu_\alpha=\mu_c\approx 0.68M_0$. }}
\end{figure}
 
\subsection{General case: $\mu_I\ne 0$, $\mu\ne 0$}

It is clear that to find the complete phase portrait of the model in
terms of the external chemical potential parameters
$\nu\equiv\mu_I/2$ and $\mu$ (at $T=0$),  one should investigate the
global minimum point (GMP) of the physical TDP $\Omega^{\rm{phys}}
(M,b,\Delta)$ (\ref{25}) vs the dynamical variables $M,b,\Delta$.
\footnote{As in the case with $b=0$, in the inhomogeneous case we did not
found local minima of the TDP (\ref{25}) of the form $(M\ne 0,\Delta\ne 0)$. }
However, in the case under consideration the problem is simplified due to the effective
reduction of external parameters. Indeed, the structure of
$\Omega^{\rm{phys}} (M,b,\Delta)$ is such that it can be considered
as a function of three dynamical variables $M,\Delta,\tilde\nu\equiv
b+\nu$ and only one external parameter $\mu$, i.e.
$\Omega^{\rm{phys}} (M,b,\Delta)\equiv F(M,\Delta,\tilde\nu;\mu)$.
So, the searching of the GMP of this function consists effectively
of two stages. First, one can find the extremum of this function
over $M$ and $\Delta$ (taking into account the results of  section
III.B) and then, as it was done in the previous section IV.A, one minimizes
the obtained expression over the variable $\tilde\nu$.
Properties of the found GMP supply the following phase structure.

If $\mu>\mu_\alpha=\mu_c\approx 0.68M_0$, then for arbitrary values
of $\nu$ we have found phases with spatially inhomogeneous
condensates, which are realized at least in the form of chiral
density waves or chiral spirals. The gap $\Delta$ is equal to zero
for these phases. The amplitude $M$ of these CDWs depends only on $\mu$
and is equal to the quantity $M(\mu)$ (see Fig. 3). However, the
chiral density wave vector $b$ depends both on $\mu$ and $\nu$,
namely
\begin{eqnarray}
b=b_0(\mu)-\nu, \label{31}
\end{eqnarray}
where the quantity $b_0(\mu)$ is also presented in Fig. 3. In the
$(\nu,\mu)$-plane (see Fig. 7) we divide this region into two CDW
phases. In the CDW$_1$  region we have the wave vector $b>0$, i.e.
here we have a clockwise twisted chiral spiral. In contrast, in the
CDW$_2$ region one obtains for chiral density waves the
counterclockwise twisted chiral spirals, since here $b<0$. For all
points of the line L of this figure, which is defined by the
relation L=$\{(\nu,\mu):\nu=b_0(\mu)\}$, the wave vector $b$ is
equal to zero. So, the points of the curve L correspond to the
homogeneous phase, where only chiral symmetry is spontaneously
broken down and the dynamical quark mass is equal to $M(\mu)\not =0$
(hence, on the line L the spatial translational invariance of the
system remains intact). Note, that the phase L is nothing else  than
the residue of the homogeneous phase 2 of Fig. 1 if the spatial
inhomogeneity of chiral condensates is taken into account. To
underline this fact, we use in Fig. 7 the notation $\mu_\alpha$,
which corresponds to the minimum point $\alpha$ of the homogeneous
phase 2 of Fig. 1, for the critical curve between the CDW and
charged pion condensation (PC) phases. However, $\mu_\alpha$
coincides with the critical value $\mu_c$ of the case $\mu_I=0$ (see
section IV.A).

As in the particular case with $\mu_I=0$ (see the previous section),
$u$-quarks are gapless excitations and $d$-quarks are gapped ones of
the CDW$_{1,2}$ phases at $\mu_I\ne 0$. The same is true for the
homogeneous phase L.

Below the line $\mu=\mu_\alpha$ of Fig. 7 the homogeneous PC phase
is arranged, since for all points of this region the GMP of the TDP
(\ref{25}) has the form $M=0,\Delta =M_0, b=0$. In this phase the
isospin $U_{I_3}(1)$ symmetry of the model is broken spontaneously.
The exception is the interval $0<\mu<\mu_\alpha$ of the $\mu$-axis,
where chiral symmetry is broken down and quarks have the mass $M_0$.

Note, both for the case of spatially homogeneous and inhomogeneous chiral condensate the 
isospin density $n_I$ in the PC phase is equal to $\nu/\pi$. Starting from the 
$\Omega^{\rm{phys}} (M,b,\Delta)$ (\ref{25}), it is possible to find the expression of this 
TDP in the CDW$_{1,2}$ phases (it is simply the expression (33) shifted by $(-\nu^2/\pi)$, in which 
$M,b$ should be replaced by $M(\mu),b_0(\mu)$, correspondingly) 
and then to calculate their isospin density $n_I=-\partial\Omega^{\rm{phys}}/\partial\mu_I$.
It turns out that in the CDW phases the isospin density is the same as in the PC phase, i.e. $n_I=\nu/\pi$.
Hence, as it is easily seen from (\ref{31}), at fixed values of $\mu$ the wave vector of 
chiral spirals is tightly (linearly) connected with isospin density. In contrast, 
in the $U_L(1)\times U_R(1)$ symmetric NJL$_2$ model without isospin chemical potential 
$\mu_I$ the wave vector $b$ shifts effectively the quark number 
chemical potential $\mu$ \cite{ohwa,thies}. Due to this reason, the quark number density $n_q$ is equal to 
$\mu/\pi$ in the CDW phase of this model. Moreover, the wave vector $b$ in this phase is proportional to $n_q$.

\section{CDW phases at nonzero temperatures}

In the case of spatially homogeneous condensates the influence of
nonzero temperature on the phase structure of the $SU_L(2)\times
SU_R(2)$ symmetric NJL$_2$ model (1) with two chemical potentials
$\mu$ and $\nu\equiv\mu_I/2$ was considered in \cite{ek2}.  Now let
us study the influence of  temperature $T$ on the phase structure of
this model in the case of an inhomogeneous chiral condensate of the
form (\ref{6}). In this case, to get the corresponding
(unrenormalized) thermodynamic potential
$\Omega_{\scriptscriptstyle{T}}(M,b,\Delta)$ one can simply start
from the expression for the TDP at zero temperature (\ref{12}) and
perform the following standard replacements:
\begin{eqnarray}
\int_{-\infty}^{\infty}\frac{dp_0}{2\pi}\big (\cdots\big )\to
iT\sum_{n=-\infty}^{\infty}\big (\cdots\big ),~~~~p_{0}\to
p_{0n}\equiv i\omega_n \equiv i\pi T(2n+1),~~~n=0,\pm 1, \pm 2,...,
\label{190}
\end{eqnarray}
i.e. the $p_0$-integration should be replaced by the summation over
an infinite set of Matsubara frequencies $\omega_n$. Summing over
Matsubara frequencies in the obtained expression  (the corresponding
technique is presented, e.g., in \cite{jacobs}), one can find for
the TDP:
\begin{eqnarray}
\Omega_{\scriptscriptstyle{T}}(M,b,\Delta)\!
&=&\frac{M^2+\Delta^2}{4G}-\int_{-\infty}^{\infty}\frac{dp_1}{2\pi}
\Big\{E^+_{\Delta}+E^-_{\Delta}+T\ln\big [1+e^{-\beta
(E^+_{\Delta}-\mu)}\big ]+T\ln\big [1+e^{-\beta
(E^+_{\Delta}+\mu)}\big ]\nonumber\\ &+&T\ln\big [ 1+e^{-\beta
(E^-_{\Delta}-\mu)}\big ]+T\ln\big [1+e^{-\beta
(E^-_{\Delta}+\mu)}\big ]\Big\}, \label{1202}
\end{eqnarray}
where $\beta=1/T$ and $E^\pm_{\Delta}$ are given in (\ref{13}).
Clearly, only the first two terms (which are the same as in the zero
temperature case), in the braces of this expression are responsible
for an ultraviolet divergency of the whole TDP (\ref{1202}). So,
regularizing the TDP (\ref{1202}) in the way as it was done in
(\ref{26}) for zero temperature TDP and then replacing $G\to
G(\Lambda)$ (see formula (\ref{16})), we can obtain in the limit
$\Lambda\to\infty$ a finite expression denoted as
$\Omega^{\rm{phys}}_{\scriptscriptstyle{T}}(M,b,\Delta)$. It is an
evident generalization of the TDP $\Omega^{\rm{phys}}(M,b,\Delta)$
(\ref{25}) to the case of nonzero temperature.  Numerical
investigations show that all possible local minima of the obtained
TDP $\Omega^{\rm{phys}}_{\scriptscriptstyle{T}}(M,b,\Delta)$ are
located in the planes $M=0$ or $\Delta=0$. So it is sufficient to
deal with corresponding restrictions of the TDP on these planes,
i.e. with the following functions,
\begin{eqnarray}
\Omega^{\rm{phys}}_{\scriptscriptstyle{T}}(M=0,b,\Delta)\!
&=&V_0(\Delta)
-\frac{2T}{\pi}\int_{0}^{\infty}dp_1\ln \Big\{\left [1+e^{-\beta
({\cal E}-\mu)}\right ]\left [1+e^{-\beta
({\cal E}+\mu)}\right ]\Big\},\label{1203}\\
\Omega^{\rm{phys}}_{\scriptscriptstyle{T}}(M,b,\Delta=0)\!
&=&V_0(M)-\frac{(\nu+b)^2}{\pi}-\frac{T}{\pi}\int_{0}^{\infty}dp_1\ln
\Big\{\left [1+e^{-\beta (E+\nu+b-\mu)}\right ]\left [1+e^{-\beta
(E+\nu+b+\mu)}\right ]\Big\}\nonumber\\
&&~~~~~~~~~-\frac{T}{\pi}\int_{0}^{\infty}dp_1\ln \Big\{\left
[1+e^{-\beta (E-\nu-b-\mu)}\right ]\left [1+e^{-\beta
(E-\nu-b+\mu)}\right ]\Big\},\label{1204}
\end{eqnarray}
where the effective potential $V_0(x)$ is given in (\ref{17}),
$E=\sqrt{p_1^2+M^2}$, and ${\cal E}=\sqrt{p_1^2+\Delta^2}$.
Comparing the global minima of the functions (\ref{1203}) and
(\ref{1204}), it is possible to establish the global minimum point
of the renormalized TDP
$\Omega^{\rm{phys}}_{\scriptscriptstyle{T}}(M,b,\Delta)$. Then, the
dependence of the global minimum point vs $T,\mu,\nu$ defines the
phase structure of the model.

Using this prescription in our numerical investigations of the TDPs
(\ref{1203})-(\ref{1204}), we have found the two $(\mu,T)$-phase
portraits of the initial NJL$_2$ model (1) depicted in Figs. 8, 9
for qualitatively different fixed values of the isospin chemical
potentials, $0\le\mu_I<2\nu_\alpha$ and $2\nu_\alpha<\mu_I$,
respectively ($\nu_\alpha\approx 0.6M_0$ is the $\nu$-coordinate of
the point $\alpha$ of Fig. 1). Note, there is a {\it first order
phase transition} on the boundaries between CDW$_{1,2}$ and homogeneous
PC or chiral symmetry breaking phases of these figures. However,
other boundaries of the phases of Figs. 8, 9 correspond to critical
curves of the second order. It is interesting to remark that for $0<\nu<\nu_\alpha$
($\nu_\alpha<\nu$) all critical curves of Fig. 8 (Fig. 9) do not depend on $\nu$.

Finally, let us take $\mu_I=0$ and compare the thermodynamical
properties of our (1+1)-dimensional NJL model (1) (see the phase
portrait of Fig. 8 at $\nu=0$) with the corresponding massless
(3+1)-dimensional NJL model with chiral $SU_L(2)\times SU_R(2)$
symmetry \cite{nickel}. It turns out that in the four-dimensional
spacetime, in contrast to the (1+1)-dimensional case, a {\it second
order phase transition}  from a homogeneous chirally broken phase to
an inhomogeneous one takes place. Moreover, depending on the value
of the dynamical quark mass in vacuum, the inhomogeneous phase in
the (3+1)-dimensional NJL model may occupy both a finite (compact)
and infinite (noncompact) region of the $(\mu,T)$-phase diagram,
whereas in our two-dimensional NJL model (1) an inhomogeneous phase
appears as a noncompact region (see Fig. 8).
\begin{figure}
 \includegraphics[width=0.45\textwidth]{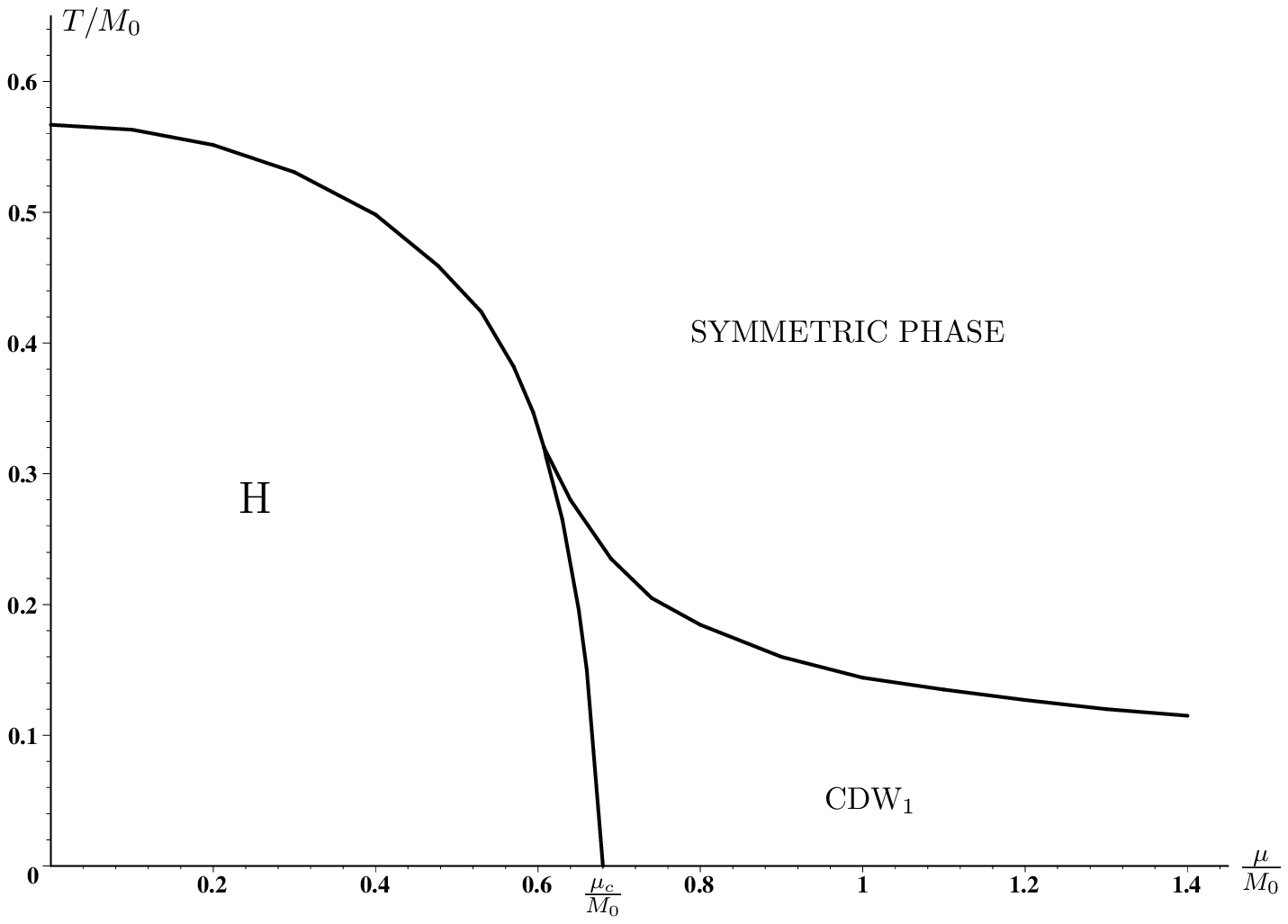}
 \hfill
 \includegraphics[width=0.45\textwidth]{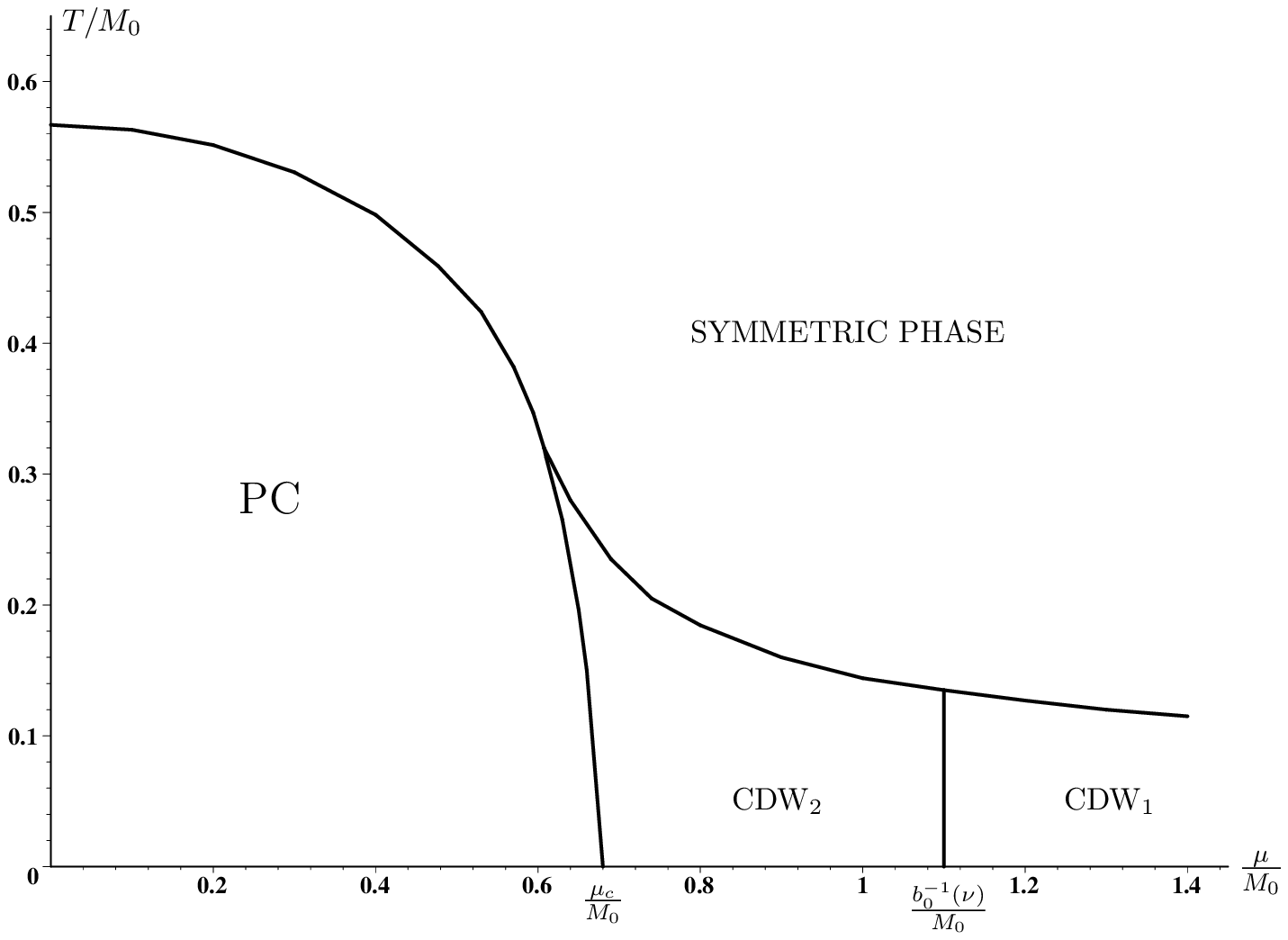}\\
\parbox[t]{0.45\textwidth}{
\caption{The $(\mu,T)$  phase portrait of the model at 
fixed $\nu$, where $0\le\nu<\nu_\alpha\approx 0.6M_0$. There, at $\nu=0$ H
denotes the homogeneous chiral symmetry breaking phase with
$M=M_0,b=0,\Delta=0$. At $0<\nu<\nu_\alpha$ H denotes the
homogeneous charged pion condensation phase (PC) with
$M=0,b=0,\Delta=M_0$. $\mu_c=\mu_\alpha\approx 0.68M_0$. 
In the symmetric phase $M=0,b=0,\Delta=0$. CDW$_{1}$ 
denotes an inhomogeneous chiral density wave phase with $b>0$.
All critical curves do not depend on $\nu$.} \label{fig:3} } \hfill
\parbox[t]{0.45\textwidth}{
\caption{The $(\mu,T)$  phase portrait at fixed $\nu$, where
$\nu_\alpha<\nu$. Here $b^{-1}_0(\nu)$ is the function inverse to
$b_0(\mu)$ defined in Fig. 3. PC denotes the homogeneous charged pion condensation phase with
$M=0,b=0,\Delta=M_0$. CDW$_{2}$ denotes the inhomogeneous chiral
density wave phase with $b<0$. All critical curves do not depend on
$\nu$. Other notations are the same as in the previous figures.} \label{fig:4} }
\end{figure}

\section{Conclusions}

It is well known that at nonzero baryon densities there might exist
phases with a spatially inhomogeneous chiral condensate. This fact
was established in the framework  of both two-dimensional GN- or
NJL-type models \cite{ohwa,thies,misha} and four-dimensional
NJL-type models \cite{3+1,nakano,nickel,maedan,zfk}, where phases
with a crystalline chiral condensate or with a CDW spatial
inhomogeneity were proved to exist at nonzero values of the baryon
chemical potential. Since the isotopic asymmetry of dense quark
matter is an inevitable reality, a more adequate investigation of
dense quark matter demands to include into consideration both
baryon, $\mu$, and isospin, $\mu_I$, chemical potentials. In this
paper and in contrast to previous papers \cite{ohwa,thies,misha}, we
study the possibility of spatially inhomogeneous chiral condensates
in the $SU_L(2)\times SU_R(2)$ symmetric NJL$_2$ model (1) including
the two above-mentioned chemical potentials in the leading order of
the large-$N_c$ expansion. For simplicity, the spatial inhomogeneity
in our consideration is realized in the form of so-called chiral
density waves or chiral spirals.

First, we have proven that at $\mu_I=0$ and $T=0$ the inhomogeneous
CDW phase is realized in this $SU_L(2)\times SU_R(2)$ symmetric
NJL$_2$ model only at sufficiently large values of $\mu$, i.e. at
$\mu>\mu_c\approx 0.68M_0$ (here $M_0$ is the dynamical quark mass
in the vacuum, i.e. at zero values of chemical potentials). In
contrast, it is well-known that in the NJL$_2$ model with continuous
$U_L(1)\times U_R(1)$ chiral symmetry the CDW phase appears at
arbitrary nonzero values of $\mu>0$ \cite{ohwa,thies}.
Moreover, it turns out that at $\mu_I=0$ the Fermi surfaces of
up/down quarks in the CDW phase are split by $2b_0(\mu)$, where
$b_0(\mu)$ is the wave vector in this phase.

Second, if $\mu_I\not = 0$ and $T=0$ then in the $(\mu_I,\mu)$ phase
diagram (see Fig. 7) the spatially inhomogeneous phases are allowed
at $\mu>\mu_\alpha=\mu_c$ and arbitrary values of $\mu_I$. This region is
divided by the curve L into two domains. In one of them each CDW is
a clockwise twisted chiral spiral, in the other -- it is a
counterclockwise twisted spiral. The amplitude of chiral density
waves does not depend on $\mu_I$. The dependence of its wave vector
$b$ on $\mu$ and $\mu_I$ is defined by the formula (\ref{31}).
Since the isospin density $n_I$ in these phases is equal to $\nu/\pi$, we see that the wave vector $b$ is linearly
connected with $n_I$. In contrast, in the $U_L(1)\times U_R(1)$-symmetric NJL$_2$ model the wave vector of the CDW phase is proportional to a quark number density \cite{ohwa,thies}.
The points of the curve L correspond to the spatially
homogeneous phase (since here $b=0$) with spontaneous chiral
symmetry breaking. Indeed, the phase L is the residue of the
homogeneous massive chirally nonsymmetric phase 2 of Fig. 1 which
shrinks to L after taking into account inhomogeneity phenomena.
Below the line $\mu=\mu_\alpha$ the homogeneous charged pion condensation phase is realized.

It turns out that at arbitrary $\mu_I$-values in all above mentioned
inhomogeneous CDW phases as well as in the L phase $u$-quarks are
gapless excitations, but $d$-quarks are gapped ones.

 Third, we have studied the influence of temperature on
the formation of the CDW phases. In particular, it was shown that at
$\mu_I=0$ the $(\mu,T)$-phase diagrams of the $SU_L(2)\times
SU_R(2)$- and $U_L(1)\times U_R(1)$ symmetric NJL$_2$ models are
quite different. Indeed, as it was proved in \cite{thies}, in the
second model the CDW phase occupies in this diagram an infinite
strip which includes points with arbitrary small $\mu$-values,
whereas in the first model (see Fig. 8) the upper boundary of this
phase is a monotonically decreasing function of $\mu$. In addition,
for rather small values of $\mu$ the CDW phase is forbidden in the
framework of the $SU_L(2)\times SU_R(2)$ symmetric NJL$_2$ model.

We finally note that in this paper we have suggested a homogeneous
pion condensate.  It would be interesting to study in future the
possibility of the spatially inhomogeneous pion condensation phase.

\subsection*{Acknowledgments}
One of the authors (V.Ch.Zh.) is grateful to Professor
M. Muller-Preussker for kind hospitality during his stay in the
particle theory group at the Institute of Physics of
Humboldt-University, where part of this work has been done, and also
to {\it Deutscher Akademischer Austauschdienst (DAAD)} for financial support.
Financial support of two other authors (N.V.G. and S.G.K.) by
Leonhard Euler stipendia of the DAAD is also appreciated.

\end{document}